\documentclass[journal,draftclsnofoot,onecolumn]{IEEEtran}

\usepackage{setspace}
\usepackage{amsmath}
\usepackage{amssymb}
\usepackage{mathrsfs}
\usepackage{amsthm}
\usepackage{multirow}
\usepackage{color}
\usepackage{longtable}
\usepackage{array}
\usepackage{url}
\usepackage{comment}
\usepackage{enumerate}
\usepackage{eucal}
\usepackage{graphics}
\usepackage{bbm}

\usepackage{algorithm}
\usepackage{algpseudocode}
\usepackage{cite}
\usepackage[table]{xcolor}

\usepackage{stmaryrd}

\usepackage{bm}
\usepackage{mathtools}
\usepackage{xparse}
\DeclarePairedDelimiterX{\set}[1]{\{}{\}}{\setargs{#1}}
\NewDocumentCommand{\setargs}{>{\SplitArgument{1}{;}}m}
{\setargsaux#1}
\NewDocumentCommand{\setargsaux}{mm}
{\IfNoValueTF{#2}{#1} {#1\,\delimsize|\,\mathopen{}#2}}

\DeclarePairedDelimiter\abs{\lvert}{\rvert}
\DeclarePairedDelimiter\ceil{\lceil}{\rceil}
\DeclarePairedDelimiter\floor{\lfloor}{\rfloor}
\DeclarePairedDelimiter\parenv{\lparen}{\rparen}
\DeclarePairedDelimiter\sparenv{\lbrack}{\rbrack}

\newcommand{\cA}{\mathcal{A}}
\newcommand{\cB}{\mathcal{B}}
\newcommand{\cC}{\mathcal{C}}
\newcommand{\cD}{\mathcal{D}}

\newcommand{\cS}{\mathcal{S}}

\newcommand{\cX}{\mathcal{X}}

\renewcommand{\leq}{\leqslant}

\renewcommand{\geq}{\geqslant}

\newcommand{\enc}{{\rm enc}}
\newcommand{\encA}{{\rm enc}_{\cA}}
\newcommand{\enci}{{\rm enc}_{i}}

\newcommand{\Hash}{{\rm Hash}}

\newcommand{\poly}{{\rm poly}}

\theoremstyle{plain}
\newtheorem{theorem}{Theorem}
\newtheorem{corollary}[theorem]{Corollary}
\newtheorem{lemma}[theorem]{Lemma}
\newtheorem{proposition}[theorem]{Proposition}

\theoremstyle{definition}

\newtheorem*{remark}{Remark}

\newcommand{\F}{\mathbb{F}}
\newcommand{\R}{\mathbb{R}}
\newcommand{\Z}{\mathbb{Z}}
\newcommand{\N}{\mathbb{N}}
\newcommand{\bT}{\mathbb{T}}
\newcommand{\bS}{\mathbb{S}}

\newcommand{\bL}{\mathbb{L}}
\newcommand{\bLM}{\mathbb{LM}}
\newcommand{\bD}{\mathbb{D}}

\newcommand{\A}{\mathtt{A}}
\newcommand{\T}{\mathtt{T}}
\newcommand{\C}{\mathtt{C}}
\newcommand{\G}{\mathtt{G}}

\newcommand{\sS}{\mathcal{S}}

\newcommand{\bpsi}{\bm{\psi}}
\newcommand{\bchi}{\bm{\chi}}
\newcommand{\beps}{\bm{\epsilon}}

\newcommand{\ve}{\mathbf{e}}
\newcommand{\vm}{\mathbf{m}}
\newcommand{\vr}{\mathbf{r}}
\newcommand{\vs}{\mathbf{s}}

\newcommand{\vu}{\mathbf{u}}
\newcommand{\vx}{\mathbf{x}}
\newcommand{\vy}{\mathbf{y}}

\newcommand{\va}{\mathbf{a}}
\newcommand{\vb}{\mathbf{b}}
\newcommand{\vc}{\mathbf{c}}
\newcommand{\vd}{\mathbf{d}}

\newcommand{\Zero}{{\mathbf{0}}}
\newcommand{\One}{{\mathbf{1}}}
\newcommand{\bbone}{\mathbbm{1}}

\newcommand{\kp}{k_+}
\newcommand{\km}{k_-}

\newcommand{\eqdef}{\triangleq}

\newcommand{\ppmod}[1]{~({\rm mod~}#1)}

\title{Improved Coding over Sets for DNA-Based Data Storage}

\author{
  Hengjia Wei and Moshe Schwartz,~\IEEEmembership{Senior Member,~IEEE}%
  \thanks{This work was supported in part by the Israel Science Foundation (ISF) under Grant 270/18.}%
  \thanks{Hengjia Wei is with the School
    of Electrical and Computer Engineering, Ben-Gurion University of the Negev,
    Beer Sheva 8410501, Israel
    (e-mail: hjwei05@gmail.com).}%
  \thanks{Moshe Schwartz is with the School
    of Electrical and Computer Engineering, Ben-Gurion University of the Negev,
    Beer Sheva 8410501, Israel
    (e-mail: schwartz@ee.bgu.ac.il).}%
}

\begin{document}

\maketitle

\begin{abstract}
  Error-correcting codes over sets, with applications to DNA storage,
  are studied. The DNA-storage channel receives a set of sequences,
  and produces a corrupted version of the set, including sequence
  loss, symbol substitution, symbol insertion/deletion, and
  limited-magnitude errors in symbols. Various parameter regimes are
  studied. New bounds on code parameters are provided, which improve
  upon known bounds. New codes are constructed, at times matching the
  bounds up to lower-order terms or small constant factors.
\end{abstract}


\section{Introduction}

\IEEEPARstart{D}{ue} to recent developments in DNA sequencing and
synthesis technologies, storing data in DNA strands has gained a lot
of interest in recent years.  One notable feature of DNA-based storage
is its ultrahigh storage densities of $10^{15}$--$10^{20}$ bytes per
gram of DNA, as demonstrated in recent experiments
(see~\cite[Table~1]{Yazdi2017.portable}). Additionally, the DNA strand
is easy to maintain and stable over millennia. These features make the
DNA strand an ideal medium to store massive amounts of data.

DNA strands can be treated as sequences comprising of four nucleotides
$\A$, $\T$, $\G$, and $\C$. In order to produce or read the strands
with an acceptable error rate, the lengths of the synthetic DNA
strands cannot be too long, usually hundreds of nucleotides. Thus, the
data in a DNA storage system is stored as a set of relatively short
strands, each of which holds a fraction of the whole data.  These
short DNA strands are dissolved inside a solution and do not preserve
the order in which they were stored. The goal of the sequencer is to
read these strands and reconstruct the data without knowledge of the
order of the sequences, even in the presence of errors.

The unordered manner of data storing in DNA storage systems motivates
the study of \emph{coding problem over sets}, following several papers
on this
topic~\cite{Lenzetal2020,Simaetal2019,SongCaiImmink,KovacevicTan,Lenzetal2019ISIT,Simaetal2020RI,Yazdietal2018,Cheeetal2020}. In~\cite{Lenzetal2020},
the authors studied the storage model where the errors are a
combination of loss of sequences, as well as symbol errors inside the
sequences, such as insertions, deletions, and substitutions. Some
lower and upper bounds were derived on the cardinality of optimal
error-correcting codes that are suitable for this model. Several
explicit code constructions are also proposed for various error
regimes. Later, \cite{Simaetal2019,SongCaiImmink} adapted the model
of~\cite{Lenzetal2020}.  In~\cite{Simaetal2019}, it was assumed that
no sequences are lost and a given number of symbol substitutions
occur.  Codes which have logarithmic redundancy in both the number of
sequences and the length of the sequences have been proposed therein.
In~\cite{SongCaiImmink}, a new metric was introduced to establish a
uniform framework to combat both sequence loss and symbol
substitutions, and Singleton-like and Plotkin-like bounds on the
cardinality of optimal codes were derived. A related model was
discussed in~\cite{KovacevicTan}, where unordered multisets are
received and errors are counted by sequences, no matter how many
symbol errors occur inside the sequences.
\cite{Lenzetal2019ISIT,Simaetal2020RI} discussed the indexing
technique to deal with the unordered nature of DNA storage.
Additionally, codes that can be used as primer addresses were proposed
in~\cite{Yazdietal2018,Cheeetal2020} to equip the DNA storage system
with random-access capabilities.

In this paper, we continue the study of coding over sets. We follow
the model of~\cite{Lenzetal2020} and present improved bounds and
constructions. We also extend the error model to include
limited-magnitude errors, following the recent application presented
in~\cite{Jainetal2020ISIT}. Our main contributions are:
\begin{enumerate}
\item We derive some new lower bounds on the redundancy of codes which
  can protect against \emph{substitutions} or \emph{deletions}. These
  results, together with some existence results, demonstrate that
  correcting deletions requires fewer redundancy bits than correcting
  substitutions. Note that a similar observation was made
  in~\cite{Lenzetal2020}, but only in the regime where there is no
  sequence loss and only a single symbol error occurs, whereas our
  results are proved for two broad parameter ranges.
\item We propose several \emph{explicit} constructions of codes having
  redundancy that is logarithmic in the number of sequences $M$,
  whereas the corresponding explicit constructions
  in~\cite{Lenzetal2020,Lenzetal2019ISIT} require $\Theta(M^c)$ bits
  of redundancy with $c>0$ a constant number.
\item We also study another error model, where data is represented by
  vectors of integers that may suffer from \emph{limited-magnitude
    errors} in some of their entries. This model is motivated by a
  recently proposed method of encoding information in DNA sequences
  which can optimize the amount of information bits per synthesis time
  unit~\cite{Jainetal2020ISIT}.  We utilize our explicit code
  constructions for substitutions to combat limited-magnitude errors.
\end{enumerate}
A summary of the bounds and constructions appearing in this paper, and
a comparison with previous results, is given in
Table~\ref{tab:summary-1} and Table~\ref{tab:summary-2}.

The remainder of the paper is organized as follows. In
Section~\ref{sec:prelim} we provide the notation and definitions used
throughout the paper. In Section~\ref{sec:corcode-bullet} we consider
channels with a fixed number of lost sequences, and a fixed number of
erroneous sequences. Section~\ref{sec:corcode-0-t} studies
codes for a channel with no sequence loss. We then study codes when
the errors are of limited magnitude in Section~\ref{sec:LMerrors}.

\section{Preliminaries}
\label{sec:prelim}

For a positive integer $n\in\N$, let $[n]$ denote the set
$\set{1,2,\ldots, n}$. For $q\in\N$, we use $\Sigma_q$ to denote a
finite alphabet with $q$ elements, $\Z_q$ to denote the cyclic group
of integers with addition modulo $q$, and $\F_q$ to denote the finite
field of size $q$. Throughout the paper, we denote the base-$q$
logarithm of a real number $a\in\R$ by $\log_q a$, and we omit the
subscript if $q=2$.

For a sequence $\va = (a_1, \ldots, a_n) \in \Sigma_q^n$, let $\va[i]$
denote the $i$th symbol $a_i$ and $\va[i,j]$ denote the subword of
$\va$ starting at position $i$ and ending at position $j$.  We use
$\abs{\va}$ to denote the length of $\va$. For two sequences $\va$ and
$\vb$, we use $(\va, \vb)$ to denote the concatenation of $\va$ and
$\vb$.  Fix an ordering of the sequences of $\Sigma_q^n$. Then every
size-$M$ subset $S \subseteq \Sigma_q^n$ can be represented by a
binary vector $\bbone(S)$, termed the \emph{characteristic vector}, of
length $q^n$ and weight $M$, where each non-zero entry indicates that
the corresponding element is contained in $S$.

\subsection{The DNA Storage Channel}

In a DNA-based data storage system, data is stored as an (unordered)
set
\[S=\set*{\vx_1,\vx_2,\ldots, \vx_M} \subseteq \Sigma_q^L\]
of $M$ distinct sequences $\vx_i$, $i \in [M]$. In practice, the
length of the sequences $L$ is in the order of a few hundreds, while
$M$ is significantly larger. A summary of typical values of $L$ and
$M$ can be found in \cite[Table I]{Lenzetal2020}. In general, we
assume that $M=q^{\beta L}$ for some $0<\beta<1$. For the sake of
simplicity, we further assume $\beta L$, i.e., $\log_q M$, is an
integer. Otherwise, a floor or ceiling is to be used in certain
places, making notation cumbersome and changing nothing in the
asymptotic analysis.

We study the \emph{$(s,t,\varepsilon)_{\bT}$-DNA storage channel
  model} defined in \cite{Lenzetal2020}. In this channel, the
sequences in $S$ are drawn arbitrarily and sequenced, possibly with
symbol errors, and we have the following assumptions on the errors:
\begin{enumerate}
\item the maximum number of sequences never drawn is $s$;
\item the maximum number of  sequences with errors  is $t$;
\item each sequence suffers at most $\varepsilon$ errors of type $\bT$.
\end{enumerate}
Note that erroneous sequences are not necessarily distinct from each
other or from the correct sequences, and that would result in sequence
losses. Thus, the output of the channel is a subset $S'$ of at least
$M-s-t$ sequences of $S$ with $t$ (or fewer) sequences each suffering
$\varepsilon$ (or fewer) errors of type $\bT$.

In~\cite{Lenzetal2020}, the authors mainly discuss the following types
of errors: substitutions $(\bS)$, deletions $(\bD)$, and a combination
of substitutions, deletions and insertions $(\bL)$, that is, $\bT\in
\set{\bS,\bD,\bL}$.  In this paper, apart from the errors mentioned
above, we also discuss \emph{limited-magnitude errors} ($\bLM$), the
model of which will be described and explained in the next subsection.

Denote
\[\cX_M^{q,L} \eqdef\set*{S\subseteq \Sigma_q^L: \abs{S}=M}.\]
For each $S \in \cX_M^{q,L}$, the \emph{error ball}
$B_{s,t,\varepsilon}^{\bT}(S)$ is defined to be the set of all
possible received $S'$ with $S$ being the input of the
$(s,t,\varepsilon)_{\bT}$-DNA storage channel. We say a subset $\cS
\subseteq \cX_M^{q,L}$ is an
\emph{$(s,t,\varepsilon)_{\bT}$-correcting code} if for any distinct
$S_1,S_2\in \cS$, it always holds that
\[ B_{s,t,\varepsilon}^{\bT} (S_1)  \cap B_{s,t,\varepsilon}^{\bT} (S_2)=\varnothing. \]
When $\varepsilon=L$,  such a code is also called an  \emph{$(s,t,\bullet)_{\bT}$-correcting code}. The \emph{redundancy} of the code $\cS$ is defined to be
\[\log_q \abs{\cX_M^{q,L}}-\log_q \abs{\cS}=\log_q \binom{q^L}{M} -\log_q \abs{\cS}. \]

In Section~\ref{sec:corcode-bullet}, we study
$(s,t,\bullet)_\bT$-correcting codes with $\bT\in \set{\bS,\bD,\bL}$,
and in
Section~\ref{sec:corcode-0-t} we study
$(0,t,\varepsilon)_\bT$-correcting codes with $\bT\in \set{\bS,\bD}$. Our results are presented in the binary case, and they
can be easily generalized to the quaternary case, i.e., $\Sigma =
\set{\A,\T,\C,\G}$.  Table~\ref{tab:summary-1} summarizes the lower
bounds and upper bounds on the redundancy of the optimal codes, while
Table~\ref{tab:summary-2} summarizes our explicit code
constructions.  These two tables also include the corresponding results
from \cite{Lenzetal2020,Lenzetal2019ISIT} for comparison.  From
Table~\ref{tab:summary-1}, we have the following observations:
\begin{enumerate}
\item For the redundancy of the optimal $(s,t,\bullet)_\bT$-correcting
  code, the lower bound almost attains the upper bound.
\item For the redundancy of the optimal
  $(0,t,\varepsilon)_\bT$-correcting code, the lower bound is nearly
  half as much as the upper bound.
\item For the sets of parameters $(s,t,\bullet)$ or
  $(0,1,\varepsilon)$, correcting deletions requires fewer redundancy
  bits than correcting substitutions.
\end{enumerate}

\begin{table*}
 \caption{Lower and upper bounds on the redundancy of optimal $(s,t,\varepsilon)_\bT$-correcting codes. Low order terms are omitted.}
  \label{tab:summary-1}
{\small
  {\renewcommand{\arraystretch}{1.5}
  \begin{tabular}{cccccccc}
    \hline\hline
    Channel    & Previous lower bound & Ref. & Imp. lower bound & Ref. & Upper bound & Ref.  \\
    \hline
    $(s,t,\bullet)_{\bL}$   & $(s+t)L+t\log M$ & \cite[Cor. 1]{Lenzetal2020} & $(s+2t)L$ & Cor.~\ref{thm:upboundpackingS}   & $(s+2t)L$ & \cite[Const. 2]{Lenzetal2020} \\
     $(s,t,\bullet)_{\bS}$   & & &  $(s+2t)L$ & Cor.~\ref{thm:upboundpackingS}   & $(s+2t)L$ & \cite[Const. 2]{Lenzetal2020} \\
      $(s,t,\bullet)_{\bD}$   &&   &  $(s+t)L$ & Cor.~\ref{thm:upboundpackingD}   & $(s+t)L$ & \cite[Const. 2]{Lenzetal2020} \\
      \hline
      $(0,t,\varepsilon)_{\bS}$   &  $t\log M+t\varepsilon \log L$  & \cite[Thm.~7]{Lenzetal2020} &   &    & $2t\log M+2t\varepsilon \log L$ & \cite[Thm. 3]{Lenzetal2020} \\
        \multirow{2}*{$(0,t,\varepsilon)_{\bD}$}   &    \multirow{2}*{$t\varepsilon \log L$} &  \multirow{2}*{\cite[Thm.~9]{Lenzetal2020}}  &   \multirow{2}*{$\floor{t/2}\log M$} &  \multirow{2}*{Thm.~\ref{thm:lowerboundD-epsilon}}   & $t\log M +2t\varepsilon \log (L/2)$ &  \cite[Thm. 4]{Lenzetal2020}\\
        & & & & & $4\varepsilon \log L$ when $t=1$ &  Thm.~\ref{thm:cons-epsilon-D} \\
    \hline\hline
  \end{tabular}
  }
  }

\end{table*}

\begin{table*}
\begin{center}
    \caption{Redundancy of the   code constructions. Low order terms are omitted. The symbol $*$ means  explicit encoding for the corresponding construction is unknown.  In the first row, $\bT\in \set{\bS,\bD,\bL}$, and $\delta= s+2t$ if $\bT\in \set{\bL,\bS}$; $\delta= s+t$ if $\bT=\bD$. In the second row, $r_o$ denotes the redundancy of an $(s,0,0)_{\bT}$-correcting code of $\cX_M^{2,L_o}$, while $r_{\bT}$ denotes the redundancy of a block-code of dimension $L_o$ and length $L$ that can correct $s$ errors of type $\bT$.}
  \label{tab:summary-2}
{
\small
  {\renewcommand{\arraystretch}{1.5}
  \begin{tabular}{ccccc}
    \hline\hline
    Channel    & Previous Construction & Ref. & Imp. Construction & Ref.    \\
    \hline
    \multirow{3}*{$(s,t,\bullet)_{\bT}$}   & $M \log e +\delta (L-\ceil{\log M})$ &  \cite[Const. 1]{Lenzetal2020}  & \multirow{3}*{{$\delta L+ 4\delta \log \log M$}} & \multirow{3}*{Cor.~\ref{cor:setcode-1}}  \\
    ~ & $\delta L$ &  \cite[Const. 2]{Lenzetal2020}*  &  ~ & ~ \\
    ~ & $\frac{1-c}{2}M^c \log M+\delta M^{1-c}(L-\log M)$ &  \cite[Const. 3]{Lenzetal2020}  &  ~ & ~ \\
    \hline
    $(s,M-s,\varepsilon)_{\bS}$  & $Mr_{\bS}+r_o$ &  \cite[Const. 4]{Lenzetal2020} & $M\varepsilon(\ceil{\log L} +1) +  sL+ 4s \log \log M$  &  Cor.~\ref{cor:cons-s-Mms-S} \\
    $(s,M-s,\varepsilon)_{\bD}$ &  $Mr_{\bD}+r_o$ & \cite[Const. 4]{Lenzetal2020} &  $4M\varepsilon \log L  + sL+o(M\log L)$ & Cor.~\ref{cor:cons-s-Mms-D}  \\
    \hline
    $(0,t,\varepsilon)_{\bS}$ &  $M\log e +4t\log M+2t\varepsilon \log L$ &  \cite[Thm. 2]{Lenzetal2019ISIT}  & $(8t+2) \log M +2t\varepsilon \ceil{\log L}+8t\log\log M $ & Cor.~\ref{cor:cons-epsilon-S}  \\
     $(0,1,\varepsilon)_{\bD}$ & $\log L$ for $\varepsilon=1$ &   \cite[Const. 5]{Lenzetal2020}   & $4\varepsilon \log L$  for any given $\varepsilon$ & Thm.~\ref{thm:cons-epsilon-D}*  \\
    \hline\hline
  \end{tabular}
  }
  }
  \end{center}
\end{table*}

\subsection{Limited-Magnitude Error Model}
Recently, a new inexpensive enzymatic method of DNA synthesis was
proposed in~\cite{LeeKalGoeBolChur19}. Unlike other synthesis methods
that focus on the synthesis of a precise DNA sequence, this method
focuses on the synthesis of runs of homopolymeric bases. Specifically,
the synthesis process proceeds in rounds. Assume at the beginning of
the round, the current string is $\vu\in \Sigma^*$. A letter $a \in
\Sigma$ is chosen, which differs from the last letter of $\vu$. A
chemical reaction is then allowed to occur for a duration of $T\in \N$
time units. The resulting string at the end of the round is
$(\vu,\underbrace{a,a,\ldots,a}_\ell)$, where $\ell$ is a random
variable whose distribution depends on the new letter being appended,
the last letter of the string at the beginning of the round, and the
duration of the chemical reaction.

For the sake of simplicity, in this paper, we consider the binary case
and  assume that the last letter of the initiator is
  $0$.  Since long runs may affect the DNA molecule's stability,
 the encoder refrains from using runs that are too
  long. Let $q$ denote the length of the longest run used by the
  encoder. Thus, every binary sequence produced by $n$ rounds of
synthesis process can be represented by a sequence
$\vr=(r_1,r_2,\ldots,r_n)$ of $\Z_q^n$, where
$r_i$ represents the length of the run appended in the
$i$th round.

Based on this enzymatic method of DNA synthesis, a new method of
encoding information in DNA strands is
proposed~\cite{Jainetal2020ISIT}.  In this method, the data is encoded
to a set of $M$ sequences $\vr_i$ of $\Z_q^L$. Then binary sequences
$\vu_i$ are produced by $L$ rounds of synthesis process described above so that
by controlling the chemical reaction, the run lengths of $\vu_i$ are
the components of {$\vr_i$}. {In the system, what we store are these  sequences $\vu_i$, whereas  the data is represented by the run-lengths of these sequences, i.e., $\set{\vr_1,\vr_2,\ldots,\vr_M}$.}

The chemical reaction may end up
shorter or longer than planned, usually by a limited amount, due to
variability in the molecule-synthesis process. Consequently, the
sequence of the run lengths of $\vu_i$ is  {$\vr_i+\ve$}, where
$\ve=(e_1,e_2,\ldots, e_L) \in [-\km,\kp]^L$ for some non-negative
integers $\kp,\km$. We say $\varepsilon$ errors that are
\emph{$(\kp,\km)$-limited-magnitude errors} $(\bLM)$ occurred, if
exactly $\varepsilon$ of the entries of $\ve$ are non-zero. This kind
of errors can also be found in other applications, like high-density
recording~\cite{KuzVin93,LevVin93} and flash
memories~\cite{CasSchBohBru10}, and the conventional coding problem to
protect against such errors has been extensively researched, e.g.,
see~\cite{WeiWangSch2020} and the references therein.

In this paper, we consider coding over sets in the presence of
limited-magnitude errors. {In this model, the codeword is still a subset $S ={
\set*{ \vr_1,\vr_2,\ldots,\vr_M}  } \subseteq \Z_q^L$. However,  each  sequence $\vr_i$ in $S$
 represents the run-lengths of a sequence $\vu_i$  produced by $L$ rounds of synthesis process. We note that these  synthesized sequences $\vu_i$'s have the same number of runs, i.e. $L$, but  may have various  lengths.
With a codeword $S  \subseteq \Z_q^L$ as input,} the
\emph{$(s,t,\varepsilon,\kp,\km)_{\bLM}$-DNA storage channel} outputs
a subsets $S'$ of $S$ with $s$ (or fewer) sequences lost and $t$ (or
fewer) sequences being corrupted by at most $\varepsilon$
$(\kp,\km)$-limited-magnitude errors. The corresponding
error-correcting code is called an
\emph{$(s,t,\varepsilon,\kp,\km)_{\bLM}$-correcting code}. { In
Section~\ref{sec:LMerrors}, we propose a construction for such
codes, which is
based on $(0,t,\varepsilon)_\bS$-correcting codes. Some bounds on the redundancy are also derived.
As before, the \emph{redundancy} of a code $\cS\subseteq \set{S \subseteq \Z_q^L; \abs{S}=M }$ is defined to be
\[\log_q \binom{q^L}{M} -\log_q \abs{\cS}, \]
where $\binom{q^L}{M}$ is the maximum  number of messages encoded by a set of $M$ sequences that are synthesized by $L$ rounds of process. We emphasize that in this model the channel receives as input $M$ vectors of length $L$ each, representing synthesis instructions for $L$ rounds. The redundancy is measured in this space. However, inside the channel, these synthesis instructions are turned into DNA sequences. These may be of different lengths for two reasons: first, the sum of run-lengths may not be equal in all the vectors. Second, the noisy synthesis process may result in different run-lengths from those intended.
}

\subsection{Some Useful Codes}
Our constructions use the well-known Reed-Solomon codes and BCH codes
as input (e.g., see~\cite{MacSlo78}). In addition, we also require the
following codes.

\begin{lemma}\cite[Theorem 1]{Simaetal2020systematic}\label{thm:deletioncode}
For any sequence $\vc \in \set{0,1}^n$ and a fixed positive integer
$\varepsilon$, there exists a hash function $\Hash_\varepsilon:
\set{0,1}^n \to \set{0,1}^{h_\varepsilon}$ with
$h_\varepsilon=4\varepsilon\log n+o(\log n)$, computable in
$O(n^{2\varepsilon+1})$ time, such that
\[\set*{(\vc, \Hash_\varepsilon(\vc)); \vc\in \set{0,1}^n}\]
forms an $\varepsilon$-deletion-correcting code. The decoding
complexity of the code is $O(n^{\varepsilon+1})$.
\end{lemma}

\begin{lemma}\label{thm:BCH-shorten}
  Let $\ell, \delta$ be positive integers such that $\delta \leq
  \frac{2^\ell-1}{\ell+1}$. Then there is a map $\enc: \F_2^{2^\ell}
  \to \F_2^{r}$ with $r\leq \delta \ell +\delta$ such that the set
  \[\set*{ (\vm,\enc(\vm)); \vm \in \F_2^{2^\ell}  } \]
  is a code of minimum Hamming distance $2\delta+1$.
\end{lemma}

\begin{IEEEproof}
  Let $n=2^{\ell+1}-1$.  Since $\delta \leq
  \frac{2^{\ell}-1}{\ell+1}$, there is a binary $[n,n-\delta (\ell+1),
    2\delta+1]$ BCH code. We may shorten this code and rearrange its
  coordinates to obtain a systematic $[2^\ell+\delta (\ell+1), 2^\ell,
    2\delta+1]$ code, and then the conclusion follows.
\end{IEEEproof}

\section{$(s,t,\bullet)_{\bT}$-Correcting Codes}
\label{sec:corcode-bullet}

In this section, we study $(s,t,\bullet)_{\bT}$-correcting codes,
where $\bT \in \set{\bS,\bD,\bL}$. We give an improved lower bound on
the redundancy of such codes, which asymptotically agrees with the
upper bound in~\cite[Theorem 13]{Lenzetal2020} up to low-order
terms. Then we give an explicit construction of codes whose redundancy
is close to this bound.

{
\subsection{Bounds Based on Constant-Weight Codes}
Fix an ordering of the vectors of  $\set{0,1}^L$. For a  subset $S$ of $\set{0,1}^L$, its characteristic
  vector, denoted $\bbone(S)$,  is a binary vector of length $2^L$  where each symbol `$1$' indicates that a specific vector is contained in the set $S$. In this way, a code $\cS \subseteq \cX_M^{2,L}$ can be represented by a binary constant-weight code
  \[\cC(\cS) \eqdef \set{\bbone(S); S\in \cS},  \]
  where all the codewords have weight $M$.

The following  result establishes the equivalence of an $(s,t,\bullet)_\bS$-correcting code and a constant-weight code of certain minimum distance.

\begin{proposition}\label{prop:storagecode2CWC}
Let $s$ and $t$ be positive integers such that $s+t\leq M$.
A code $\cS \subseteq \cX_M^{2,L}$ is an $(s,t,\bullet)_\bS$-correcting code if and only if the corresponding constant-weight code $\cC(\cS)$ has minimum Hamming distance at least $2(s+2t)+2$.
\end{proposition}

\begin{IEEEproof}
  Denote $\delta\eqdef s+2t$. We first show that if $\cC(\cS)$ has
  minimum distance $\geq 2\delta+2$ then $\cS$ is an
  $(s,t,\bullet)_\bS$-correcting code.  Note that for a codeword $S
  \in \cS$, the $s$ deletions and $t$ substitutions of sequences in
  $S$ correspond to at most $s+t$ asymmetric $1\to 0$ errors and $t$
  asymmetric $0 \to 1$ errors in $\bbone(S)$. Thus, if $\cC(\cS)$ has
  minimum Hamming distance $2\delta+2$, we can correct these $\delta$
  substitution errors in $\bbone(S)$, and then recover $S$. That is,
  $\cS$ is an $(s,t,\bullet)_\bS$-correcting code

  In the other direction, if $\cC(\cS)$ has minimum Hamming distance
  less than $2\delta+2$, then it is at most $2\delta$ since the
  distance between two sequences of the same weight is even. Let
  $\bbone(S)$ and $\bbone(S')$ be two codewords in $\cC(\cS)$ with
  distance at most $2\delta$. Necessarily, $S$ and $S'$ share at least
  $M-\delta$ sequences.  W.l.o.g., we may assume that
  \[ S= \set*{\vu_1,\vu_2,\ldots, \vu_{M-\delta}, \va_1,\ldots,\va_s, \vb_1,\ldots, \vb_{2 t}} \in \cS\]
  and
  \[S'= \set*{\vu_1,\vu_2,\ldots, \vu_{M-\delta}, \va'_1,\ldots,\va'_s, \vb'_1,\ldots, \vb'_{2 t}} \in \cS.\]

  Note that in the $(s,t,\bullet)_\bS$-DNA storage channel, the result
  of an erroneous sequence could be any sequence of
  $\set*{0,1}^L$. Hence, after going through the channel, both $S$ and
  $S'$ can generate the set
  \[ \set*{\vu_1,\vu_2,\ldots, \vu_{M-\delta}, \vb_1,\vb_2,\ldots,\vb_t, \vb'_{t+1},\ldots, \vb'_{2 t}}.\]
  This implies that $\cS$ is not an $(s,t,\bullet)_\bS$-correcting
  code.
\end{IEEEproof}

The following upper bound on the size of a constant-weight code can be found in \cite{GraSlo80}.

\begin{lemma}\label{lem:upperboundCWC}
Let $\cC$ be a binary constant-weight code of length $n$, weight $w$ and minimum distance $2\tau$. Then
\[\abs{\cC} \leq \frac{\binom{n}{w-\tau+1}}{\binom{w}{w-\tau+1}}.\]
\end{lemma}

}

\begin{corollary}\label{thm:upboundpackingS}
  Let $s$ and $t$ be positive integers such that $s+t\leq M$. For any
  $(s,t,\bullet)_\bS$-correcting code $\cS$, the code size satisfies
  \[\abs{\cS } \leq  \frac{ \binom{2^L}{M-s-2t}  }{ \binom{M}{M-s-2t} }.\]
  In particular, if both $s$ and $t$ are fixed, the redundancy of an
  $(s,t,\bullet)_\bS$-correcting code is at least
  \[ (s+2t) L -\log ((s+2t)!) -o(1).\]
\end{corollary}

\begin{IEEEproof} { The first bound is obtained directly from Proposition~\ref{prop:storagecode2CWC} and Lemma~\ref{lem:upperboundCWC}.}
  If $s$ and $t$ are both fixed, then $\delta$ is fixed, and the
  redundancy satisfies
  \begin{align*}
    \log \binom{2^L}{M} -\log \abs{\cS}  & \geq \log \frac{  \binom{2^L}{M} \binom{M}{M-\delta}   }{ \binom{2^L}{M-\delta }  }   = \log \frac{(2^L-M+1)(2^L-M+2)\cdots (2^L-M+\delta)  }{ \delta!}   \\
    & \geq  \delta \log (2^L-M+1)-\log (\delta!)= \delta L-\delta \log \frac{2^L}{2^L-M+1} -\log (\delta!)\\
    & =  \delta L-\delta \parenv*{\log e} \ln \parenv*{1+\frac{M-1}{2^L-M+1}} -\log (\delta!)\\
    & \overset{(a)}{\geq} \delta L -\frac{(M-1)\delta \log e}{2^L-M+1} -\log (\delta!) \\
    &\overset{(b)}{=}\delta L -\log (\delta !) -o(1),
  \end{align*}
  where $(a)$ holds as $\ln (1+x) \leq x$ for all $x>-1$, and $(b)$
  holds as $M=2^{\beta L}$ for some constant $\beta <1$.
\end{IEEEproof}

\begin{remark}
  Note that an $(s,t,\bullet)_{\bL}$-correcting code is also an
  $(s,t,\bullet)_{\bS}$-correcting code. According to
  Corollary~\ref{thm:upboundpackingS}, the redundancy of an
  $(s,t,\bullet)_{\bL}$-correcting code is at least \[(s+2t) L -\log
  ((s+2t) !) -o(1),\] which improves upon the bound $(s+t)L+t\log
  M+o(1)$ in \cite[Corollary~1]{Lenzetal2020} since $L>\log M$.
  Moreover, this new bound is asymptotically tight since there exist
  $(s,t,\bullet)_{\bL}$-correcting codes of redundancy $(s+2t)L$
  \cite[Construction~2 and Theorem~13]{Lenzetal2020} .
\end{remark}

Using the same argument, we have the following results for codes that
can correct deletions.
{
\begin{proposition}\label{prop:storagecode2CWC-deletion}
Let $s$ and $t$ be positive integers such that $s+t\leq M$.
A code $\cS \subseteq \cX_M^{2,L}$ is an $(s,t,\bullet)_\bD$-correcting code if and only if the corresponding constant-weight code $\cC(\cS)$ has minimum Hamming distance $2(s+t)+2$.
\end{proposition}}

\begin{corollary}\label{thm:upboundpackingD}
  Let $s$ and $t$ be positive integers such that $s+t\leq M$. For any
  $(s,t,\bullet)_\bD$-correcting code $\cS$, the code size satisfies
  \[\abs{\cS } \leq  \frac{ \binom{2^L}{M-s-t}  }{ \binom{M}{M-s-t} }.\]
  In particular, if both $s$ and $t$ are fixed, the redundancy of an
  $(s,t,\bullet)_\bD$-correcting code is at least
  \[ (s+t) L -\log ((s+t) !) -o(1).\]
\end{corollary}

Note that  the $(s,t,\bullet)_\bD$-correcting code in  \cite[Table 2]{Lenzetal2020} has  redundancy $(s+t)L$. This redundancy
almost meets the lower bound in Corollary~\ref{thm:upboundpackingD}, and
is strictly less than the minimum redundancy $(s+2t) L -\log ((s+2t)
!) -o(1)$ required for correcting substitutions.

\subsection{Explicit Code Constructions}

For $(s,t,\bullet)_{\bT}$-correcting codes, three constructions can be
found in \cite{Lenzetal2020}. In particular,
\cite[Construction~1]{Lenzetal2020} and
\cite[Construction~3]{Lenzetal2020} can produce codes with redundancy
$\Theta(M)$ and $\Theta(M^c\log M)$ for some real constant $c>0$,
respectively, while \cite[Construction~2]{Lenzetal2020} requires
$\delta L$ bits of redundancy, where
\begin{equation}\label{def:delta}
\begin{split}
  \delta = & \begin{cases}
    s+2t, & \text{if $\bT\in \{\bS, \bL$\},} \\
    s+t, & \text{ if $\bT= \bD$.}
\end{cases}
\end{split}
\end{equation}
 Noting that $ L= \beta^{-1}\log M$, the latter
construction is much better than the former two. However, efficient
encoding\footnote{By ``efficient", we mean the complexity of the encoding  is in $\poly(M)$ time.} for \cite[Construction~2]{Lenzetal2020} is unknown.

In this section, we propose an explicit construction of
$(s,t,\bullet)_{\bT}$-correcting codes with redundancy at most
{$\delta L + O(\log \log M)$}, that can be encoded efficiently {when $\delta$ is fixed}.  Our method modifies
\cite[Construction~1]{Lenzetal2020}, where the code contains the
codewords $S=\set{\vx_1,\vx_2\ldots,\vx_M} \subseteq \F_2^L$ such that
\begin{enumerate}
\item $\vx_i = (I(i), \vu_i)$ for $1\leq i \leq M$, where $I(i)$ is
  the binary representation of $i-1$;
\item if each $\vu_i$ is regarded as an element of $\F_{2^{L-\log
    M}}$, the sequence $(\vu_1,\vu_2,\ldots,\vu_M)$ belongs to a given
  $[M,M-\delta, \delta+1]$ MDS code over $\F_{2^{L-\log M}}$, where
  $\delta\eqdef s+2t$.
\end{enumerate}

In our construction, instead of using the binary representations
$I(i)$ to index the sequences in the codeword $S$, we use sequences of
length $L'$ with $L' >\log M$ to index those
sequences. Specifically, let $\log M < L' <L$, and let $\cA$ be the
collection of all the subsets of $\F_2^{L'}$ of size $M$.  Each set
$A=\set{\va_1,\va_2,\ldots,\va_M}\in \cA$ is regarded as a set of
addresses\footnote{Throughout this paper we keep enumerating the
  sequences in the address set $A$ in a descending lexicographic
  order.}. For each codeword $S$ of the proposed DNA-storage code, we
associate an address set $A\in\cA$ to $S$ and use the addresses
$\va_i$'s to index the sequences in $S$. It is worth noting that, in
our construction, different codewords may be associated with different
address sets, while in \cite[Construction~1]{Lenzetal2020} all the
codewords use the same set of addresses, i.e., $\set*{I(i);1\leq i\leq
  M}$.

{
  Besides $\cA$, our construction also requires the following codes:
  \begin{itemize}
  \item A binary systematic
$[2^{L'}+\delta({L'}+1),2^{L'},2\delta+1]$ code $\cC_{\cA}$ from Lemma~\ref{thm:BCH-shorten}. For each $A\in \cA$, let
  $\encA(A)$ be the vector of $\F_2^{\delta(L'+1)}$ such that
  $(\bbone(A),\encA(A))\in\cC_{\cA}$.
  \item A hash function $\Hash_\delta:
  \set{0,1}^{\delta(L'+1)} \to \set{0,1}^{h}$ from
  in Lemma~\ref{thm:deletioncode}, where  $h=4\delta \log (L')+o(\log L').$
  \item An $[M, M-\delta, \delta+1]$ MDS code  $\cB$ over $\F_{2^{L-L'}}$. Such a code exists whenever $L-L'\geq \log M$.
  \end{itemize}

\begin{theorem}\label{th:con-infepsilon} Let $\bT \in \set{\bS, \bD,\bL}$.
  Given $s,t,M$ and $L$, let $\delta$ be defined as in \eqref{def:delta} and let $L'$ be a positive integer such that $\log
  M< \min \set{L',L-L'}$. Suppose that $\delta(L'+1)+h\leq M-\delta$,
  and assume $\cA$, $\cC_{\cA}$, $\Hash_\delta$, and $\cB$, are as above.


  Denote by $\cS$ the collection of the sets $\set{\vx_i = (\va_i,
    \vu_i);1\leq i\leq M} \subseteq \F_2^{L}$ that satisfy all of
  the following conditions:
  \begin{enumerate}
  \item
    \label{item:c1}
    $A\eqdef\set{\va_1,\ldots,\va_M} \in \cA$ (indexed
    lexicographically).
  \item
    \label{item:c2}
    $(\vu_1[1], \vu_2[1], \ldots,
    \vu_{\delta(L'+1)+h}[1])=(\encA(A),\Hash_\delta(\encA(A)))$.
  \item
    \label{item:c3}
    $(\vu_1,\vu_2,\ldots,\vu_M) \in \cB$, where $\vu_i$ is
    treated as an element of $\F_{2^{L-L'}}$.
  \end{enumerate}
  Then the code $\cS$ is an $(s,t,\bullet)_{\bT}$-correcting code of size
  \[\abs*{\cS}=\binom{2^{L'}}{M} 2^{(L-L')(M-\delta)-\delta(L'+1)-h}.\]
\end{theorem}

\begin{figure}[t]
  \centering
  \includegraphics[width=10cm, trim={0 0 0 0}, clip]{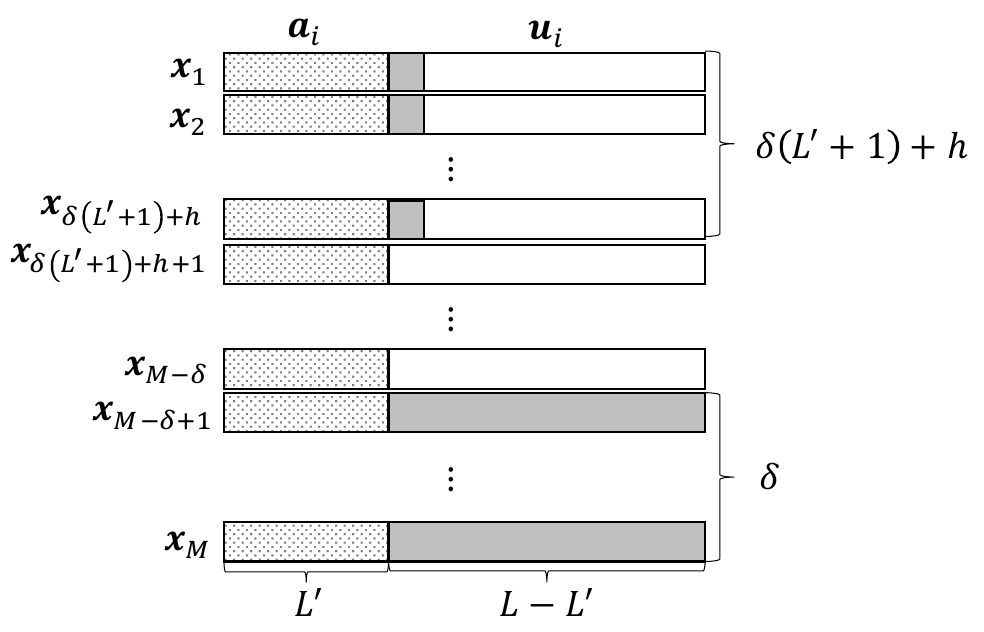}
  \caption{The construction in Theorem~\ref{th:con-infepsilon}. The dotted parts represent the addresses, the blank parts represent the information bits, and the gray parts represent the check bits of the codeword.}
  \label{fig:con1}
\end{figure}

\begin{IEEEproof}
    We first check the size of $\cS$, i.e., the number of possible
    choices of $\set{\vx_1, \vx_2,\ldots, \vx_M}$ that satisfy all the
    conditions above.  The construction is depicted in
    Fig.~\ref{fig:con1}.  From~\ref{item:c1}), there are $\binom{2^{L'}}{M}$
    choices of $A=\set{\va_1,\va_2,\ldots,\va_M}$. Given $A$,
    according to~\ref{item:c2}), there are $2^{L-L'-1}$ choices of $\vu_i$ for
    each $1\leq i\leq \delta(L'+1)+h$, and $2^{L-L'}$ choices of
    $\vu_i$ for each $\delta(L'+1)+h +1\leq i\leq M-\delta$.  Finally,
    for $M-\delta+1 \leq i\leq M$, according to~\ref{item:c3}) these $\vu_i$'s
    are determined by $(\vu_1, \ldots, \vu_{M-\delta})$ since the code
    $\cB$ has dimension $M-\delta$. Thus, the size of $\cS$ stated in
    the theorem is correct.

    Now, we show that $\cS$ is an $(s,t,\bullet)_{\bT}$-correcting
    code by describing a decoding procedure. Suppose that the input of
    the channel is a set $S=\set*{\vx_i=(\va_i,\vu_i); 1\leq i\leq M}$
    and the output is a set $S'=\set*{\vx_i'=(\va_i',\vu_i'); 1\leq
      i\leq M-s'}$, where $0\leq s'\leq s$ and the sequences in $\cS$
    and $\cS'$ are enumerated in a descending lexicographic order. Our
    decoding has the following steps:

    \textbf{Step 1:} Let $\vc'\eqdef (\vu_1'[1], \vu_2'[1], \ldots,
    \vu_{\delta L'+h-s}'[1])$. Then $\vc'$ can be obtained from $\vc
    \eqdef (\vu_1[1], \vu_2[1], \ldots, \vu_{\delta(L'+1)+h}[1])$ by
    deleting $s'$ elements and inserting $t'$ elements for some $s'$
    and $t'$. In the following, we give an upper bound on $s'+t'$.  We
    treat the channel as having two stages. In the first stage, only
    $s$ sequences are lost but no surviving sequence is
    corrupted. Hence, in this stage there are $s$ deletions in
    $\vc$. In the second stage, at most $t$ sequences are
    corrupted. Note that each erroneous sequence results in either one
    substitution or one deletion and one insertion. In total, we have
    that $s'+t'\leq s+2t$. If $\bT=\bS$ or $\bL$, then $s'+t'\leq
    s+2t= \delta$. Due to the error-correcting capability of the
    deletion code in Lemma~\ref{thm:deletioncode}, we are able to
    recover $\vc$ from $\vc'$ \footnote{{We note that the
        decoding in Lemma~\ref{thm:deletioncode} is designed to
        correct deletions, while we need to correct both deletions and
        insertions. However, since the number of codewords is
        $2^{\delta(L'+1)}$, if $L'=O(\log M)$, we still can decode in
        $\poly(M)$ time.}}, and so also $\encA(A)$.  If $\bT=\bD$, we
    further delete $\vu_\ell'[1]$ from $\vc'$ if the length of
    $\vx_\ell'$ is less than $L$ and denote the resulting sequence as
    $\vc''$. Note that in this case each erroneous sequence only
    results in one deletion. Thus, $\vc''$ can be obtained from $\vc$
    by deleting at most $s+t=\delta$ elements. Hence, we can recover
    $\vc$ from $\vc''$, and so also $\encA(A)$.

    \textbf{Step 2:} Let
    \[A'\eqdef\set*{\va'; \vx'=(\va',\vu')\in S', \abs{\vx'}=L}.\]
    Here we ignore the multiplicity of the elements in $A'$, namely,
    $A'$ is a simple set. Since at most $s$ sequences of $S$ are lost
    and $t$ sequences are erroneous, the characteristic vector
    $\bbone(A')$ can be treated as an erroneous version of $\bbone(A)$
    with at most $s+2t$ substitution errors if $\bT=\bS$ or $\bL$, or
    with at most $s+t$ substitutions if $\bT=\bD$. Thus, we can run
    the decoding algorithm of $\cC_{\cA}$ on the concatenation of
    $\bbone(A')$ and $\encA(A)$ to recover $\bbone(A)$, and so, the
    address set $A=\set*{\va_1, \va_2,\ldots, \va_M }$.

    \textbf{Step 3:} For each $1\leq i \leq M$, if there is a unique
    sequence $\vx'=(\va',\vu')\in S'$ with $\abs{\vx'}=L$ such that
    $\va'=\va_i$, let $\vu_i'=\vu'$; otherwise, let $\vu_i'$ be the
    empty string (representing an erasure). Consider the sequence
    $(\vu_1',\vu_2',\ldots,\vu_M')$ over $\F_{2^{L-L_1}}$. Using the
    same argument as that in the proof of
    \cite[Construction~1]{Lenzetal2020}, one can show that this
    sequence can be obtained from $(\vu_1,\vu_2,\ldots,\vu_M)$ by at
    most $s'$ erasures and $t'$ substitutions with $s'+2t'\leq \delta$
    if $\bT\in \set{\bS, \bL}$, or by at most $s+t$ erasures if $\bT=\bD$. So, we
    may run the decoding algorithm of $\cB$ on
    $(\vu_1',\vu_2',\ldots,\vu_M')$ to recover
    $(\vu_1,\vu_2,\ldots,\vu_M)$.
\end{IEEEproof}

We now turn to show that the redundancy of the code constructed in
Theorem~\ref{th:con-infepsilon} is  $\delta L+O(\log \log M)$.

\begin{corollary}\label{cor:setcode-1}
Let $\bT \in \set{\bS, \bD,\bL}$.
  Given $s,t,M$ and $L$, let $\delta$ be defined as in~\eqref{def:delta}.
  Assume that $M\geq 2\delta\log M +6\delta \log\log M$ and $L> 3\log M$, then
  there is an $(s,t,\bullet)_{\bT}$-correcting code of $\cX_M^{2,L}$ with
  redundancy at most
  \[\delta L+ 4\delta \log \log M+o(\log \log M).\]
\end{corollary}

\begin{IEEEproof}
  Let $L'=2\log M$. Then the conditions of Theorem~\ref{th:con-infepsilon}
  are satisfied. Applying this theorem, we get an
  $(s,t,\bullet)_{\bT}$-correcting code $\cS$ with redundancy
  \begin{align*}
    & \log \binom{2^{L}}{M} - \log\abs{\cS} \\
    = &    \log \binom{2^{L}}{M} - \log \binom{2^{L'}}{M} -(L-L')(M-\delta)+\delta(L'+1)+h\\
    \leq & M(L-L')+M\log \frac{2^{L'}}{2^{L'}-M}-(L-L')(M-\delta)+\delta(L'+1)+h\\
    \leq &\delta(L-L')+\delta(L'+1)+h+2\log e\\
    = & \delta L+ 4\delta \log \log M+o(\log \log M),
  \end{align*}
  as claimed.
\end{IEEEproof}

Next, we analyze the complexity of the encoding and the decoding of
the codes in Theorem~\ref{th:con-infepsilon}. Assume that the message
is
\[(a,\vm)\in \sparenv*{0,\binom{2^{L'}}{M}-1}\times  \F_2^{(L-L')(M-\delta)-\delta(L'+1)-h}.\]
Let us first examine the encoding process. First, we encode the
integer $a$ of $\sparenv*{0,\binom{2^{L'}}{M}-1}$ to a subset
$A=\set{\va_1,\ldots,\va_M} \in \cA$. This can be done by a greedy
algorithm in $O(M2^{L'}L')$ time. We run the encoding of $\cC_{\cA}$
on $\bbone(A)$ to obtain $\encA(A)$. Since $\cC_{\cA}$ has dimension
$2^{L'}$ and redundancy $\delta(L'+1)$, the time complexity is
$O(2^{L'}\delta(L'+1))$. Then, we compute $\Hash_\delta(\encA(A))$,
which can be done in $O((\delta L')^{2\delta+1})$ time. We write
$\encA(A)$, $\Hash_\delta(\encA(A))$ and the word $\vm$ of
$\F_2^{(L-L')(M-\delta)-\delta(L'+1)-h}$ onto $\vu_i$, where $1\leq i
\leq M-\delta$. Finally, we use the encoding of the code $\cB$ to
determine $\vu_i$, where $M-\delta+1\leq i \leq M$. The time
complexity of this step is $O(\delta(M-\delta)(L-L'))$.  In summary,
taking $L'=2\log M$, and since $\delta$ is a constant, the total time
complexity of the encoding is $O(M^3 \log M)$ operations over $\F_2$.

For the decoding, in Step~1 we first sort the sequences in the
received codeword $S'$ and then decode $\encA(A)$ from $\vc'$.  This
may be done in $O(M^2 \log M)$ time. Steps~2 and~3 simply run the
decoding of the codes $\cC_{\cA}$ and $\cB$.  Hence, the complexity of
the whole decoding is $O(M^3)$.

\subsection{Application to $(s,M-s,\varepsilon)_{\bT}$-Correcting Codes}

In \cite{Lenzetal2020}, Lenz et al. gave a concatenation
method to construct $(s,M-s,\varepsilon)_{\bT}$-correcting codes. This construction uses an $(s,0,0)_\bT$-correcting
  code as the inner code. Lenz et al. suggested to use their Constructions 1, 2 or 3 to obtain
the required $(s,0,0)_\bT$-correcting code. Now we can use the
  code from  Theorem~\ref{th:con-infepsilon} as the inner code to construct the $(s,M-s,\varepsilon)_{\bT}$-correcting code and the whole construction is still explicit.
  }

\begin{lemma}\cite[Construction~4 and Lemma~8]{Lenzetal2020}\label{thm:concatenation} Let   $\bT\in \set*{\bS,\bD}$.
  Let $\cS_o \subseteq \cX_M^{2,L_o}$ be an $(s,0,0)_\bT$-correcting
  code and $\cC_i$ be a block-code of dimension $L_o$ and length $L$ that can
  correct $\varepsilon$ errors of type $\bT$. Let $\enci(\cdot) :
  \set{0,1}^{L_o} \to \set{0,1}^L$ be an encoder of the code $\cC_i$.
  Define
  \[\cS \eqdef\set*{ S \in \cX_M^{2,L}; S= \bigcup_{\vx_o \in S_o}  \set*{\enci(\vx_o)}, S_o \in \cS_o }.\]
  Then $\cS$ is an $(s,M-s,\varepsilon)_{\bT}$-correcting code.
\end{lemma}

Let $r_o\eqdef \log {2^{L_o} \choose M}- \log \abs{\cS_o}$ and
$r_i\eqdef L-L_o$ be the redundancy of the outer code and the inner
code, respectively.  When $L_o\geq 2\log M$, the redundancy of the
code $\cS$ in the construction above can be bounded as follows:
\begin{align*}
  &\log {2^L \choose M} -\log \abs{\cS} \\
  = &  \log {2^L \choose M} - \log {2^{L_o} \choose M}+\log {2^{L_o} \choose M}- \log \abs{\cS_o} \\
  \leq & M(L-L_o)+M\log \frac{2^{L_o}}{2^{L_o}-M}+\log {2^{L_o} \choose M}- \log \abs{\cS_o} \\
  \leq & M r_i+r_o+2\log e.\\
\end{align*}

Using the code from Lemmas~\ref{thm:deletioncode}
or~\ref{thm:BCH-shorten} as the inner code and the code from
Corollary~\ref{cor:setcode-1} as the outer code, we obtain the following
results.

\begin{corollary}\label{cor:cons-s-Mms-S}
  For any positive integers $s,M,L_o$ with $L_o > 3\log M$ and $M\geq
  2s\log M +5s \log\log M$, and a fixed positive integer
  $\varepsilon$, there is an $(s,M-s,\varepsilon)_{\bS}$-correcting
  code $\cS \subseteq \cX_M^{2,L}$ with $L={L_o}+\varepsilon(
  \ceil{\log{L_o}}+1)$ and redundancy at most
  \[M\varepsilon(\ceil{\log L_o} +1) +  sL_o+ 4s \log \log M+o(\log \log M).\]
\end{corollary}

\begin{corollary}\label{cor:cons-s-Mms-D}
  For any positive integers $s,M,L_o$ with $L_o > 3\log M$ and $M\geq
  2s\log M +5s \log\log M$, and a fixed positive integer
  $\varepsilon$, there is an $(s,M-s,\varepsilon)_{\bD}$-correcting
  code $\cS \subseteq \cX_M^{2,L}$ with $L={L_o}+4\varepsilon
  \log{L_o} +o(\log L_o)$, and redundancy at most
  \[4M\varepsilon \log L_o  + sL_o+o(M\log L_o).\]
\end{corollary}

\section{$(0,t,\varepsilon)_{\bT}$-Correcting Codes}
\label{sec:corcode-0-t}

In this section, we study channels that have no sequence loss, namely,
$(0,t,\varepsilon)_{\bT}$-correcting codes with $\bT \in
\set{\bS,\bD}$. We improve the lower bound on the redundancy of
optimal $(0,t,\varepsilon)_{\bD}$-correcting codes and propose new
constructions for $(0,1,\varepsilon)_{\bD}$-correcting codes and
$(0,t,\varepsilon)_{\bS}$-correcting codes.

\subsection{$(0,t,\varepsilon)_{\bD}$-Correcting Codes}

Let $s=0$ and $t$ and $\varepsilon$ be fixed. Lenz et
al.~\cite{Lenzetal2020} showed the following two lower bounds on the
number of redundancy bits that are required to correct substitutions
and deletions, respectively.

\begin{lemma}\cite[Theorem 7 and Theorem 9]{Lenzetal2020}\label{thm-lowbound-fixed}
  For fixed positive integers $t$ and $\varepsilon$, the redundancy of
  a $(0,t,\varepsilon)_{\bS}$-correcting code is at least
  \[t\log M +t\varepsilon \log L +o(1),\]
  while the redundancy of a $(0,t,\varepsilon)_{\bD}$-correcting code is at least
  \[t\varepsilon \log L +o(1).\]
\end{lemma}

Since $M=2^{\beta L}$, it follows that $t\varepsilon \log L = O(\log
\log M)$. Thus Lemma~\ref{thm-lowbound-fixed} implies that in the
$(0,t,\varepsilon)_\bT$-DNA storage channel, correcting deletions may
require fewer redundancy bits than correcting deletions.  When
$t=\varepsilon=1$, Lenz et al. demonstrated this by constructing a
class of $(0,1,1)_{\bD}$-correcting codes of redundancy $\log
(L+1)$. Their method utilized the fact that one can directly identify
the unique erroneous sequence with deletions.  We generalize their
method to $\varepsilon >1$ and obtain the following result.

\begin{theorem}\label{thm:cons-epsilon-D}
  Let $M$, $L$, and $\varepsilon$, be positive integers. Let
  $\Hash_\varepsilon: \set{0,1}^L \to \set{0,1}^{h_\varepsilon}$ be
  the hash function defined in Lemma~\ref{thm:deletioncode}, where
  $h_\varepsilon=4\varepsilon \log L +o(\log L)$.  For any $\va\in
  \set{0,1}^{h_\varepsilon}$, define
  \[\cS_\va \eqdef \set*{S\in  \cX_M^{2,L} ; \sum_{\vx \in S} \Hash_\varepsilon(\vx) = \va .}\]
  Then $\cS_{\va}$ is a $(0,1,\varepsilon)_{\bD}$-correcting
  code. Furthermore, there is at least one choice of
  $\va\in\set{0,1}^{h_\varepsilon}$ such that the code $\cS_{\va}$ has
  redundancy at most
  \[4\varepsilon \log L+o(\log L).\]
\end{theorem}

\begin{IEEEproof}
  Suppose that the input of the channel is $S$ and the output is
  $S'$. W.l.o.g., assume that the deletions occur in the sequence
  $\vx_0 \in S$ and result in a sequence $\vx_0' \in S'$. We can
  identify the $M-1$ error-free sequences from $S'$ as they have
  length $L$ and the erroneous sequence $\vx_0'$ has length less than
  $L$. So we have that
  \[\Hash_{\varepsilon}(\vx_0)=\va-\sum_{\vx\in S', \abs{\vx}=L}  \Hash_{\varepsilon}(\vx).\]
  We then recover the sequence $\vx_0$ from $\vx_0'$ by running the
  decoding algorithm mentioned in Lemma~\ref{thm:deletioncode}.
\end{IEEEproof}

As a consequence, we observe that as long as $t=1$, correcting
deletions indeed requires fewer redundancy bits than correcting
substitutions. When $t\geq 2$, however, the lower bound for deletions
in Lemma~\ref{thm-lowbound-fixed} is not tight. We can improve it
exponentially, from $\Omega(\log \log M)$ to $ \floor{t/2} \log M$.

\begin{theorem}\label{thm:lowerboundD-epsilon}
  Let $M,L,t,\varepsilon$ be positive integers with $t$ and
  $\varepsilon$ fixed. Assume that $L>3\log M+\varepsilon$.  Then the
  redundancy of a $(0,t,\varepsilon)_{\bD}$-correcting code is at
  least
  \[   \floor{t/2} \log M + \floor{t/2} \varepsilon-O(1).\]
\end{theorem}

\begin{IEEEproof}
  For each $S\in \cX_M^{2,L}$, we index the sequences
  $\vx_1,\vx_2,\ldots, \vx_M$ in $S$ such that they are in a
  descending lexicographic order. Denote $L_\varepsilon\eqdef
  L-\varepsilon$. Let $S|_{L_\varepsilon}$ be the multiset projection
  of $S$ onto the first $L_\varepsilon$ bits, i.e., the multiset
  \[S|_{L_\varepsilon}\eqdef \set{\vx_1[1,L_\varepsilon], \vx_2[1,L_\varepsilon],\ldots, \vx_M[1,L_\varepsilon]}.\]

  Partition $\cX_M^{2,L}$ into equivalence classes $\cD_1, \cD_2,
  \ldots, \cD_m$ such that $S$ and $S'$ are in the same subset if and
  only if their multiset projections $S|_{L_\varepsilon}$ and
  $S'|_{L_\varepsilon}$ are the same. Each $S$ contains $M$ distinct sequences, so in each projection, each sequence of length $L_\varepsilon$ occurs at most $2^\varepsilon$ times. Thus, the number of equivalence
    classes $m$ is exactly the number of ways to throw $M$ indistinguishable
    balls into $2^{L_\varepsilon}$ distinguishable urns, each of capacity
    limited to $2^{\varepsilon}$ balls. This number is known to be (e.g., see~\cite[Ex.~6, p.~360]{Cha02})
  \[m= \sum_{j=0}^{2^{L_\varepsilon}}(-1)^j \binom{2^{L_\varepsilon}}{j}\binom{2^{L_\varepsilon}+M-j(2^{\varepsilon}+1)-1}{2^{L_{\varepsilon}}}.\]

  This expression for $m$, however, is inconvenient to work with, so
  now we give an upper bound on $m$.  W.l.o.g., we assume that for
  $1\leq i \leq m_1$, where $m_1\leq m$, the multiset projection in
  each $\cD_i$ contains $M$ different sequences of length
  $L_\varepsilon$, and for $m_1 < i \leq m$, the multiset projection
  contains fewer than $M$ distinct sequences.  For $1\leq i \leq m_1$,
  since the projection in each $\cD_i$ has $M$ different sequences,
  $\cD_i$ has size exactly $({2^\varepsilon})^{M}$, and so
  \begin{equation}\label{eq:m1}
    m_1= \binom{2^{L_\varepsilon}}{M} \leq  \frac{  \binom{2^L}{M} }{2^{\varepsilon M} }.
  \end{equation}
  The number of equivalence classes with repetitions is
  \[m-m_1\leq \sum_{K=1}^{M-1} \binom{2^{L_\varepsilon}}{K} K^{M-K},\]
  where in this expression, $K$ counts the number of distinct
  sequences in the multiset, $\binom{2^{L_\varepsilon}}{K}$ gives the
  number of choices of these distinct sequences, and $K^{M-K}$ counts
  how the remaining {$M-K$} sequences as repetitions of the $K$ distinct
  ones (we ignore the $2^{\varepsilon}$ upper limit on repetition).
  Since $L>3\log M+\varepsilon$, when $K\leq M-2$, we have
  \[
  \frac{\binom{2^{L_\varepsilon}}{K}K^{M-K}}{\binom{2^{L_\varepsilon}}{K+1}(K+1)^{M-K-1}}
  = \frac{(K+1)^2}{2^{L_\varepsilon}-K}\parenv*{\frac{K}{K+1}}^{M-K} <1.
  \]
  It follows that $\binom{2^{L_\varepsilon}}{K} K^{M-K}$ is increasing
  in $K$. Hence,
  \begin{equation}\label{eq:m2}
    m-m_1 = \sum_{K=1}^{M-1}   \binom{2^{L_\varepsilon}}{K} K^{M-K}  \leq \binom{2^{L_\varepsilon}}{M-1} M^2.
  \end{equation}
  We show that the number in \eqref{eq:m1} is larger than that in
  \eqref{eq:m2}:
  \begin{align*}
    \frac{  \binom{2^L}{M} }{2^{\varepsilon M} } \bigg/  \binom{2^{L_\varepsilon}}{M-1} M^{2}&
    = \frac{  (2^{L}-M+1)(2^{L}-M+2)(2^{L}-M+3)\cdots 2^{L}  }{M (2^{L_\varepsilon}-M+2)(2^{L_\varepsilon}-M+3)\cdots 2^{L_\varepsilon}  }
    \cdot \frac{1}{2^{\varepsilon M}M^2}\\
    & \geq \frac{  (2^{L_\varepsilon}-M+1)(2^{L_\varepsilon}-M+2)(2^{L_\varepsilon}-M+3)\cdots 2^{L_\varepsilon}  }{M^3 (2^{L_\varepsilon}-M+2)(2^{L_\varepsilon}-M+3)\cdots 2^{L_\varepsilon}  }\\
    & = \frac{ 2^{L_\varepsilon}-M+1 }{M^3} \geq 2^{L_\varepsilon-1 -3\log M}\geq 1.
  \end{align*}
  Hence,
  \[ m \leq  \frac{  \binom{2^L}{M} }{2^{\varepsilon M-1} }. \]

  Now, let $\cS$ be a $(0,t,\varepsilon)_{\bD}$-correcting
  code. According to the pigeonhole principle, there is one
  $\cD_{i_0}$, where $1\leq i_0\leq m$, such that $\cS \cap \cD_{i_0}$
  has size at least $ \frac{ \abs{\cS} }{m}$. Denote $\cS^*\eqdef \cS
  \cap \cD_{i_0}$. So
  \begin{equation} \label{eq:lb-subcodeS}
    \abs{\cS^*}\geq \frac{ \abs{\cS}  }{m} \geq   \frac{ \abs{\cS} }{ \binom{2^L}{M} \big/ 2^{\varepsilon M -1} }.
  \end{equation}

  Let $\Sigma \eqdef \set{0,1}^\varepsilon$ and
  \[ \cC \eqdef \set*{  (\vx_1[L_\varepsilon+1,L], \vx_2[L_\varepsilon+1,L], \ldots, \vx_M[L_\varepsilon+1,L]) \in \Sigma^M ; \set{\vx_1,\vx_2,\ldots, \vx_M} \in \cS^* }.\]
  We point out that while $\set{\vx_1,\vx_2,\ldots, \vx_M} \in \cS^*$
  is a set, at this point we use the lexicographic ordering to assign
  the indices, resulting in a single vector,
  $(\vx_1[L_\varepsilon+1,L], \vx_2[L_\varepsilon+1,L], \ldots,
  \vx_M[L_\varepsilon+1,L]) \in \Sigma^M$.

  We contend that $\cC\subseteq\Sigma^M$ is a code of minimum Hamming
  distance at least $t+1$; otherwise, if there are two codewords in
  $\cC$ that have a Hamming distance of at most $t$, then the two
  corresponding codewords in $\cS^*$ would be confusable in the
  $(0,t,\varepsilon)_{\bD}$-DNA storage channel { by deleting the length-$\varepsilon$ suffixes  corresponding to the positions in which the codewords in $\cC$ differ}. Hence, by using the
  Hamming bound on $\abs{\cC}$, which is the same as $\abs{\cS^*}$, we
  have that
  \begin{equation}\label{eq:ub-subcodeS}
    \abs{\cS^*}\leq \frac{2^{\varepsilon M}}{ \sum_{i=0}^{ \floor{t/2}} \binom{M}{i} (2^\varepsilon -1)^i }.
  \end{equation}
  Combining \eqref{eq:lb-subcodeS} and \eqref{eq:ub-subcodeS}, we have that
  \[ \abs{\cS}  \leq \frac{2 \binom{2^L}{M} }{ \sum_{i=0}^{ \floor{t/2}} \binom{M}{i} (2^\varepsilon -1)^i } .\]
  Hence
  \begin{align*}
    \log \binom{2^L}{M} - \log \abs{\cS} & \geq  \log \parenv*{ \sum_{i=0}^{ \floor{t/2}} \binom{M}{i} (2^\varepsilon -1)^i } -1\\
    & = \floor{t/2} \log M + \floor{t/2} \varepsilon-O(1).
  \end{align*}
\end{IEEEproof}

\begin{remark}
  When $t\geq 2$, it is still unclear whether there are
  $(0,t,\varepsilon)_{\bD}$-correcting codes of redundancy less than
  the lower bound $t\log M +t\varepsilon \log L +o(1)$ in
  Lemma~\ref{thm-lowbound-fixed} for substitutions. The
  Gilbert-Varshamov bound shows that the redundancy of optimal
  $(0,t,\varepsilon)_{\bD}$-correcting codes is at most $t\log M
  +2t\varepsilon \log (L/2)$, see \cite[Thm. 4]{Lenzetal2020}. This
  upper bound is nearly twice the improved lower bound for deletions
  in Theorem~\ref{thm:lowerboundD-epsilon}, but is still a bit larger
  than the lower bound for substitutions.
\end{remark}

\subsection{$(0,t,\varepsilon)_{\bS}$-Correcting Codes}

Next, we consider the problem of finding explicit constructions for
$(0,t,\varepsilon)_{\bS}$-correcting codes.  A related problem is
studied in \cite{Lenzetal2019ISIT}. The input of that channel is a set
$S$ of $M$ indexed sequences of $\F_2^L$, i.e., $S=\set*{(I(i),
  \vu_i); 1\leq i\leq M}\subseteq \F_2^{L}$, and at the decoder no
sequences of $S$ are lost, and at most $t$ sequences are erroneous
where each $I(i)$ suffers at most $\varepsilon_1$ substitution errors
and each $\vu_i$ at most $\varepsilon_2$ substitution errors.  The
construction proposed in \cite{Lenzetal2019ISIT} requires $M\log e +4t\log M+2t\varepsilon_2 \log L$
bits of redundancy\footnote{The redundancy in \cite{Lenzetal2019ISIT},
  which is defined to be $M(L-\log M) -\log \abs{\cS}$, is different
  from the one defined in this paper and \cite{Lenzetal2020}.}.

Our construction involves the following codes:
\begin{itemize}
\item A code
  \begin{equation}\label{eq:addresscode}
    \cA\eqdef\set*{ \set*{\va_1,\va_2,\ldots,\va_M} \subseteq \F_2^{L'} ; \va_1=\One, d_H(\va_i,\va_j) \geq 2\varepsilon+1 \textup{ for all } i\not=j },
  \end{equation}
  where $\log M <L'<L$. {The idea behind this code comes from \cite{Lenzetal2019ISIT} and the cardinality analysis may be found in \cite{Shinkaretal17}.} \cite{Simaetal2019} shows that such a code can
  be constructed using an algorithm which is similar to the
  Gilbert-Varshamov bound so that
  \[\abs{\cA} \geq \frac{ \prod_{i=2}^M  \parenv*{2^{L'}-(i-1)Q}  }  {(M-1)!},\]
  where $Q=\sum_{i=0}^{2\varepsilon} \binom{L'}{i}$ is the size of a
  Hamming ball of radius $2\varepsilon$ in
  $\F_2^{L'}$. \cite{Simaetal2020RI} proposes an efficient encoding
  algorithm in $\poly(M,L,\varepsilon)$ time.
  \item A binary $[2^{L'}+2t({L'}+1),2^{L'},4t+1]$ code $\cC_{\cA}$
    from Lemma~\ref{thm:BCH-shorten}. {For each $A\in \cA$, let
  $\encA(A)$ be the vector of $\F_2^{2t(L'+1)}$ such that
  $(\bbone(A),\encA(A))\in\cC_{\cA}$.}
    \item {A hash function $\Hash_{2t}:
  \set{0,1}^{2t(L'+1)} \to \set{0,1}^{h}$ from
  in Lemma~\ref{thm:deletioncode}, where  $h=8t \log (L')+o(\log L').$}
  \item A binary $[L-L', L-L'-r, 2\varepsilon+1]$ code {$\cC_1$} with
    $r\eqdef \varepsilon \ceil{\log (L-L')}$. To obtain such a code, we
    may shorten a binary $[n,n-r,2\varepsilon+1]$ BCH code with
    $n=2^{\ceil{\log (L-L')}}-1$.
  \item An $[{M, M-{\tilde{r}} }, 2t+1]$ code {$\cC_2$} over
    $\F_{2^r}$. Let $q=2^r$ and $m = \ceil{ \log {(M+1 )}/r}$. Then this
    code can be obtained by shortening a $[q^m-1,q^m-1-\tilde{r},
      2t+1]$ BCH code over $\F_{q}$. Note that
\[\tilde{r} <  2t m = 2t \ceil{ \log  { (M+1 )} /r}.  \]
  \end{itemize}

\begin{theorem}\label{th:con}
 Given $t,M$ and $L$,  let $L'$ be a positive integer such that $\log
  M< \min \set{L',L-L'}$.
{Let $\cA,$ $\encA(A),$ $\Hash_{2t},$ $\cC_1$ and  $\cC_2$ be defined as above. Let $H$ be the parity check matrix of $\cC_1$.
Suppose that $M \geq  2t(L'+1)+h$ where $h=8t \log (L')+o(\log L')$.}

  Denote by $\cS$ the collection of the sets $\set{\vx_1=(\va_i,\vu_i)
    ; 1\leq i\leq M} \subseteq \F_2^L$ that satisfy all of the
  following:
  \begin{enumerate}
  \item
    \label{item:d1}
    $A\eqdef\set{\va_1,\va_2,\ldots,\va_M} \in \cA$.
  \item
    \label{item:d2}
    { $(\vu_1[1], \vu_2[1], \ldots,
    \vu_{2t(L'+1)+h}[1])=(\encA(A),\Hash_{2t}(\encA(A)))$.}
  \item
    \label{item:d3}
    $( { \vs_1, \vs_2},\ldots,\vs_M)\in\cC_2$,
    where $ \vs_i\eqdef\vu_i H^T$ is regarded as an element of
    $\F_{2^r}$ \footnote{ {This kind of coding scheme is known as tensor product code, see \cite{Lenzetal2019ISIT} and the reference therein.}}.
  \end{enumerate}
  Then the code $\cS$ is a $(0,t,\varepsilon)_{\bS}$-correcting code
  of size
  \[\abs{\cS}=\abs{\cA}2^{M(L-L')-2t(L'+1){-h -r\tilde{r}}}.\]
\end{theorem}

\begin{figure*}[htbp]
  \centering
  \includegraphics[width=10cm, trim={0 0 0 0}, clip]{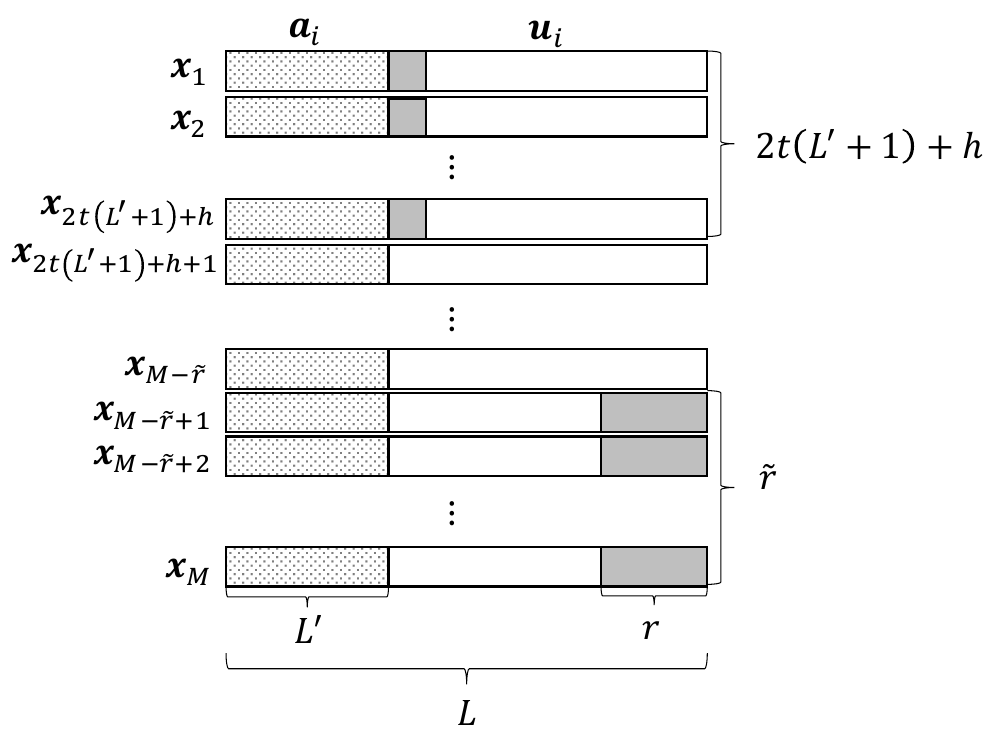}
  \vspace{-4mm}

  \caption{The construction in Theorem~\ref{th:con}. The dotted parts represent the addresses, the blank parts represent the information bits and the gray parts represent the check bits of the codeword.}
  \vspace{-3mm}

  \label{fig:con2}
\end{figure*}

\begin{IEEEproof}
  We first check the size of $\cS$, i.e., the number of possible
  choices of $\set{\vx_1, \vx_2,\ldots, \vx_M}$ satisfying all the
  conditions above.  {The construction is depicted in
    Fig.~\ref{fig:con2}.  For~\ref{item:d1}), there are $|\cA|$
    choices of $A=\set{\va_1,\va_2,\ldots,\va_M}$.  Given $A$,
    according to~\ref{item:d2}), there are $2^{L-L'-1}$ choices of
    $\vu_i$ for each $1\leq i\leq 2t(L'+1)+h$, and $2^{L-L'}$ choices
    of $\vu_i$ for each $2t(L'+1)+h +1\leq i\leq M-\tilde{r}$.
    According to~\ref{item:d3}), the sequences $\vu_i$, where $1\leq
    i\leq M-\tilde{r}$, can determine $\vs_i$ for all $1\leq i \leq
    M$, as $\cC_2$ has dimension $M-\tilde{r}$. Now, for each
    $M-\tilde{r}+1\leq i \leq M$, given $\vs_i$, there are
    $2^{L-L'-r}$ choices of $\vu_i$, since $\cC_1$ is a code of
    dimension $L-L'-r$. Thus, the size of $\cS$ is as stated.
}

  Now, we show that $\cS$ is a $(0,t,\varepsilon)_{\bS}$-correcting
  code by providing a decoding procedure. Suppose that the input of
  the channel is a set $S=\set*{\vx_i=(\va_i,\vu_i); 1\leq i\leq M}$
  and the output is { a
  set $S'=\set*{\vx_i'=(\va_i',\vu_i'); 1\leq i\leq M}$, where the sequences in $\cS$ and $\cS'$ are
  enumerated in a descending lexicographic order.} Our decoding has the
  following steps.
  \begin{enumerate}
  \item  Let $\vc'\eqdef (\vu_1'[1], \vu_2'[1], \ldots,
    \vu_{ 2t(L'+1)+h}'[1])$. Then $\vc'$ can be obtained from  $\vc = (\vu_1[1], \vu_2[1], \ldots, \vu_{2t(L'+1)+h}[1])$ by deleting $t'$
    elements and inserting $t'$ elements with $t'\leq t$. Due to the error-correcting capability of the deletion code in
    Lemma~\ref{thm:deletioncode}, we are able to recover $\vc$ from $\vc'$, and so $\encA(A)$.
  \item Denote $A'\eqdef \set*{ \va'; (\va',\vu') \in S' }$. Noting
    that there are at most $t$ sequences of $S$ suffering errors, the
    distance between $\bbone(A)$ and $\bbone(A')$ is at most
    $2t$. Thus, we may run the decoding algorithm of $\cC_{\cA}$ on
    the concatenation of $\bbone(A')$ and $\encA(A)$ to
    recover $\bbone(A)$, and so the set $A$. Denote the sequences of
    $A$ as $\va_1, \va_2,\ldots, \va_M$, in a descending lexicographic
    order. Since $A$ is a code of minimum distance $2\varepsilon+1$,
    for each $1\leq i \leq M$, there is a unique sequence of $S'$ such
    that its length-$L'$ prefix is of distance at most $\varepsilon$
    from $\va_i$. Denote this sequence as $\vx_i'=(\va_i',\vu_i')$.
  \item Compute the syndromes $\vs_i'=\vu_i'H^T$ for $1\leq i\leq
    M$. Since there are at most $t$ sequences of $S'$ are erroneous,
    we can run the decoding algorithm of $\cC_2$ on $(\vs_1',\vs_2',
    \ldots, \vs_M')$ to recover $(\vs_1,\vs_2, \ldots, \vs_M)$.
  \item For each $1\leq i\leq M$, choose an arbitrary solution $\vy_i$
    to $\vy_i H^T= \vs_i$. Then $(\vu_i-\vy_i)H^T=\vu_i H^T- \vy_i
    H^T=\Zero$, and so $\vu_i-\vy_i$ is a codeword of $\cB$. Run the
    decoding algorithm of $\cC_1$ on $\vu_i' - \vy_i$ to recover $\vu_i
    - \vy_i$, and so $\vx_i=(\va_i,\vu_i)$.
  \end{enumerate}
\end{IEEEproof}

\begin{corollary}\label{cor:cons-epsilon-S}
  If {$L\geq 4\log M +4\varepsilon^2+1$, $t$ and
    $\varepsilon$ are fixed positive integers, and $M$ is sufficiently
    large}, then there is a $(0,t,\varepsilon)_{\bS}$-correcting code
  with redundancy at most
  \[(8t+2) \log M +{2t\varepsilon \ceil{\log L}+8t\log\log M +o(\log\log M)}.\]
  \end{corollary}

\begin{IEEEproof}
  Let $L'=3\log M+4 \varepsilon^2+1$. Then $\log
  M< \min \set{L',L-L'}$ and  $M \geq  2t(L'+1)+h$, and
  Theorem~\ref{th:con} shows that there is a
  $(0,t,\varepsilon)_{\bS}$-correcting code $\cS$ of redundancy
  \begin{align*}
    &    \log \binom{ 2^L}{ M} - \log\abs{\cS}\\
    \leq &    \log \binom{ 2^L}{M} - \log \frac{ \prod_{i=2}^M  \sparenv{2^{L'}-(i-1)Q}  }  {(M-1)!}  - M(L-L') +2t(L'+1)+ {r\tilde{r}+h}\\
    \leq & ML' -\log M -    (M-1) \log \parenv*{  2^{L'}-MQ} +2t(L'+1)+{h+r\tilde{r}} \\
    = & L' -\log M + (M-1) \log \parenv*{1+\frac{MQ}{ 2^{L'}-MQ}} +2t(L'+1)+{h+r\tilde{r}} \\
    \leq & L' -\log M +\frac{(M-1)MQ}{ 2^{L'}-MQ}\log e+2t(L'+1)+{h+r\tilde{r}} \\
    \overset{(a)}{\leq}  &  L' -\log M + \log e+2t(L'+1)+{h+r\tilde{r}} \\
     \leq & (8t+2) \log M + { 2t\varepsilon \ceil{\log L}+8t\log\log M +o(\log\log M)},
  \end{align*}
  where the inequality $(a)$ holds as $Q=\sum_{i=0}^{2\varepsilon}
  \binom{L'}{i}<(L')^{2\varepsilon}$, and $M^2(L')^{2\varepsilon} \leq
  2^{L'}$ when $L'=3\log M+4 \varepsilon^2+1$, see \cite[Appendix
    E]{Simaetal2019}.
\end{IEEEproof}

\section{Codes Correcting Limited-Magnitude Errors}
\label{sec:LMerrors}

In this section, we study the limited-magnitude
errors, motivated by DNA-storage channels that involve such
errors~\cite{Jainetal2020ISIT}. Unlike substitution errors studied in
previous sections, a limited-magnitude error replaces an integer entry
$x\in\Z$ by $x'$ such that $x-\km\leq x'\leq x+\kp$. Thus, we no
longer use the binary alphabet, instead, using $\Z_q$.

{

First, we derive the following Gilbert-Varshamov bound.

\begin{lemma}\label{lm:GVbound-LM}
There is an $(s,t,\varepsilon,\kp,\km)_{\bLM}$-correcting code $\cS
  \subseteq \cX_M^{q,L}$ of size at least
  \[  \binom{q^L}{M} \bigg/ \sparenv*{\binom{M}{s} \binom{M-s}{t} \binom{M-s+t-1}{t} \binom{q^L}{s} \parenv*{\sum_{i=0}^{\varepsilon}\binom{L}{i}(\kp+\km)^i}^{2t}} .  \]
  \end{lemma}

  \begin{IEEEproof}
Given a set $S \in \cX_M^{q,L}$, let $B^{\bLM}(S)$ be the set of all possible received $S'$ with $S$ being the input of the $(s,t,\varepsilon,\kp,\km)_{\bLM}$-DNA storage channel.  We shall give an upper bound on the number of sets $\tilde{S} \in \cX_M^{q,L}$ such that $B^{\bLM}(S)\cap B^{\bLM}(\tilde{S})\not = \varnothing$.

Note that if  two balls $B^{\bLM}(S)$ and $B^{\bLM}(\tilde{S})$ intersect, necessarily in the intersection there is a set $S'$ of size at most $M-s$. The number of $S' \in B^{\bLM}(S)$ with $\abs{S'}\leq M-s$  is no more than $\binom{M}{s} \binom{M-s}{t}V^{t}$, where $V=\sum_{i=0}^{\varepsilon}\binom{L}{i}(\kp+\km)^i$.
Now, fix a set $S' \in B^{\bLM}(S)$, we count the number of $\tilde{S} \in \cX_M^{q,L}$ such that $S' \in B^{\bLM}(\tilde{S})$. First, there are at most $\binom{q^L}{s}$ choices  for the lost sequences. Next, noting that each sequence of $S'$ may come from different sequences of $\tilde{S}$ with errors, the number of the choices for the erroneous sequences of $\tilde{S}$ is at most $\binom{M-s+t-1}{t} V^t$. Hence,  the number of  $\tilde{S}$ such that $B^{\bLM}(S)\cap B^{\bLM}(\tilde{S})\not = \varnothing$ is at most
\[\binom{M}{s} \binom{M-s}{t} \binom{M-s+t-1}{t} \binom{q^L}{s}V^{2t}.\]
The conclusion follows from this estimation and a greedy argument.
\end{IEEEproof}

 \begin{corollary}\label{cor:LM-upbound-3}When $s$ and $t$ are fixed and $\varepsilon=L$, there is a $q$-ary $(s,t,\bullet,\kp,\km)_{\bLM}$-correcting code with
  redundancy at most
 \[(s+2t)\log_q M +2tL\log_q(\kp+\km+1)+sL+O(1);\]
when $s$ and $\varepsilon$ are fixed, and $t=M-s$, there is a $q$-ary $(s,M-s,\varepsilon,\kp,\km)_{\bLM}$-correcting code with
  redundancy at most
\[2M(\varepsilon \log_q L+\varepsilon \log_q (\kp+\km)+\log_q2-\log_q \varepsilon !)  +o(M);\]
when $s,t,\varepsilon$ are fixed, there is a $q$-ary  $(s,t,\varepsilon,\kp,\km)_{\bLM}$-correcting code of redundancy at most
\[(s+2t) \log_q M +sL +2t\varepsilon \log_q L +2t\varepsilon \log_q (\kp+\km) +O(1).\]
\end{corollary}


For lower bounds on the redundancy, we follow the approach in \cite{Lenzetal2020} and derive the following sphere-packing bounds. The proofs are  the same as that of \cite[Theorem 7 and Theorem 8]{Lenzetal2020}, and we omit here.

\begin{lemma}\label{lm:spherepacking-LM}
For fixed $s,t,\varepsilon$ and fixed $0<\beta<1$, any $(s,t,\varepsilon,\kp,\km)_{\bLM}$-correcting code $\cS
  \subseteq \cX_M^{q,L}$ with $M=q^{\beta L}$ has redundancy at least
\[sL+t\log_q M +t\varepsilon  (\log_q L+\log_q(\kp+\km))-\log_q(s!t!(\varepsilon !)^t)+o(1).\]
\end{lemma}

\begin{lemma}
For fixed $s,\varepsilon$ and fixed $0<\beta<1$, any $(s,M-s,\varepsilon,\kp,\km)_{\bLM}$-correcting code $\cS
  \subseteq \cX_M^{q,L}$ with $M=q^{\beta L}$   has redundancy at least
\[M\varepsilon(\log_qL+\log_q ((\kp+\km)))+O(M).\]
\end{lemma}

}

We can make use of $(0,t,\varepsilon)_{\bS}$-codes which we
constructed earlier as a basis for constructing codes for the
limited-magnitude error scenario.

\begin{theorem}\label{thm:conmod}
  Let $\cC \subseteq \cX_M^{p,L}$ be a
  $(0,t,\varepsilon)_{\bS}$-correcting code with $p \geq
  \kp+\km+1$. Suppose that for each $C\in\cC$, the minimum Hamming
  distance of $C$ is at least $2\varepsilon+1$. Define
  \[  \cS\eqdef\set*{ S=\set{\vx_i ; 1\leq i\leq M} \subseteq \Z_q^L ;  \set{\vx_i \ppmod{p};1\leq i\leq M } \in \cC }.\]
  Then $\sS$ is a $(0,t,\varepsilon,\kp,\km)_{\bLM}$-correcting code
  over $\Z_q$ of size $\abs{\cS}\geq\floor{q/p}^{ML} \abs{\cC}$.
\end{theorem}

\begin{IEEEproof}
  We prove the theorem by describing a decoding procedure. Let
  $S=\set{\vx_1,\vx_2,\ldots,\vx_M} \in \cS$ be the input of the
  channel, and $S'$ be the output. Since each codeword $C\in \cC$ has
  minimum Hamming distance $2\varepsilon+1$, we can see that $S'$
  comprises $M$ distinct sequences. W.l.o.g., we assume that $S'
  =\set{\vy_1,\vy_2,\ldots,\vy_{M}}$ and there is a permutation $\pi$
  (to be determined later) such that for all $1\leq i\leq M$, $\vx_i$
  yields $\vy_{\pi(i)}$ after passing through the channel.

  Let $\bchi_i=\vx_i \ppmod{p}$ and $\bpsi_i=\vy_i \ppmod{p}$. Then
  $\bpsi_{\pi(i)}-\bchi_i=\vy_{\pi(i)} -\vx_i \ppmod{p}$. That implies
  that $\bpsi_{\pi(i)}$ is an erroneous version of $\bchi_{i}$ with at
  most $\varepsilon$ positions being corrupted by substitution
  errors. Thus, we may run the decoding algorithm of the
  $(0,t,\varepsilon)_{\bS}$-correcting code $\cC$ on
  $\set{\bpsi_1,\bpsi_2,\ldots, \bpsi_M}$ to recover the set
  $\set{\bchi_1,\bchi_2,\ldots, \bchi_M}$.

  Now, for each $\bchi_i$, we have $ d_H(\bchi_i, \bpsi_{\pi(i)})\leq
  \varepsilon$. We claim that $d_H(\bchi_i,\bpsi_j) > \varepsilon$ for
  all $j\not = \pi(i)$. Otherwise,
  \[ d_H(\bchi_i,  \bchi_{\pi^{-1}(j)} ) \leq d_H(\bchi_i,  \bpsi_j) + d_H(\bpsi_j,  \bchi_{\pi^{-1}(j)} ) \leq 2\varepsilon,\]
  which contradicts the assumption that the minimum Hamming distance
  of $C$ is at least $2\varepsilon+1$.  Therefore, the permutation
  $\pi$ can be determined by computing the Hamming distance between
  $\bchi_i$ and $\bpsi_j$ for all $1\leq i,j\leq M$. Denote $\beps_i=
  \bpsi_{\pi(i)} - \bchi_i \ppmod{p}$ and let
  $\ve_i=(e_1^{(i)},e_2^{(i)},\ldots,e_L^{(i)})$ where
  \[
  e_\ell^{(i)} \eqdef \begin{cases} \epsilon_\ell^{(i)}, \textup{\ \ if $0\leq \epsilon_\ell^{(i)} \leq \kp$;} \\
    \epsilon_\ell^{(i)}-p, \textup{\ \ otherwise.}\\
  \end{cases}
  \]
  Then $\vx_i$ can be decoded as $\vx_i=\vy_{\pi(i)}-\ve_i$.
\end{IEEEproof}

Let \[ r_p(\cC) \eqdef \log_p \binom{p^L}{M} -\log_p \abs{\cC}.\]
Then we have that
\[ \log_q \abs{\cC}=\log_q p \log_p \abs{\cC}= \log_q \binom{p^L}{M}-r_p(\cC)\log_q p .\]
If $p \mid q$ and $\log_p M < L/2$, then the redundancy of the code $\cS$ is
\begin{align*}
 & \log_q \binom{q^L}{M} - \log_q\abs{\cS}\\
= & \log_q \binom{q^L}{M} -ML \log_q \parenv*{\frac{q}{p}} - \log_q \abs{\cC}\\
= & \log_q \binom{q^L}{M}-\log_q \binom{p^L}{M}-ML \log_q \parenv*{\frac{q}{p}}+ r_p(\cC)\log_q p \\
\leq & M\log_q\frac{p^L}{p^L-M} + r_p(\cC)\log_q p\\
\leq & \frac{M^2}{p^L-M} \log_q e+ r_p(\cC)\log_q p \\
= & r_p(\cC)\log_q p+o(1).
\end{align*}

{
We note that the code in
Corollary~\ref{cor:cons-epsilon-S} satisfies the condition in
Theorem~\ref{thm:conmod}, i.e., each codeword has minimum Hamming
distance $2\varepsilon+1$, thus we may use it as the input code and
get the following result.

}

\begin{corollary}\label{cor:LM-upbound-2}
  Let $q>0$ be an even integer. If $L\geq (2t+1)(3\log M+4
  \varepsilon^2+2)+\varepsilon \ceil{\log L}-1$, and $t$ and
  $\varepsilon$ are fixed, then there is a $q$-ary
  $(0,t,\varepsilon,1,0)_{\bLM}$-correcting code with redundancy at
  most
  \[(8t+2) \log_q M +(2t+1) \varepsilon \log_q L +O(1).\]
\end{corollary}
{
We note that this redundancy is larger than the Gilbert-Varshamov  bound  $2t\log_q M +2t\varepsilon \log_q L+O(1)$.

Next, we modify Lemma~\ref{thm:concatenation} to construct an $(s,M-s,\varepsilon, \kp,\km)_{\bLM}$-correcting code.
\begin{lemma}\label{lm:concatenation-LM}
  Let $\cS_o \subseteq \cX_M^{2,L_o}$ be an $(s,0,0)_\bT$-correcting
  code and let $\cC_i$ be a $q$-ary block-code of size $2^L_o$ and length $L$ that can
  correct $\varepsilon$ $(\kp,\km)$-limited-magnitude errors. Let $\enci(\cdot) :
  \set{0,1}^{L_o} \to \Z_q^L$ be an encoder of the inner code $\cC_i$.
  Define
  \[\cS \eqdef\set*{ \bigcup_{\vx_o \in S_o}\set*{\enci(\vx_o)}\in\cX_M^{q,L};  S_o \in \cS_o }.\]
  Then $\cS$ is an $(s,M-s,\varepsilon,\kp,\km)_{\bLM}$-correcting code.
\end{lemma}

\begin{proof}
We first use the inner code to correct all the limited-magnitude errors, and then use the outer code to recover all the lost sequences.
\end{proof}

Let $L_o>3\log M$. We may use the code in
Corollary~\ref{cor:setcode-1} as the outer code with redundancy
$r_o=sL+O(\log \log M)$.  As for the inner code, let $p$ be the
smallest prime number such that $p > \kp+\km$, and $L$ be the smallest
integer such that $\frac{q^L}{p^{L-K}}\geq 2^{L_o}$, where
$L-K=\ceil{2\varepsilon(1-1/p)\log_p L }$. We take the code in
\cite[Theorem 5 and Corollary 8]{WeiWangSch2020} of size
$\frac{q^L}{p^{L-K}}$ as the inner code. The redundancy of the resulting
code is
\begin{align*}
 & \log_q \binom{q^L}{M} - \log_q \abs{\cS_o} \\
 \leq & \log_q \binom{q^L}{M} -\log_q \binom{2^{L_o}}{M} + \log_q \binom{2^{L_o}}{M}- \log_q \abs{\cS_o} \\
 \leq & M \log_q \frac{q^L}{2^{L_o}} + M\log_q \frac{2^{L_o} }{2^{L_o}-M} + \frac{r_o}{\log q}\\
 \leq & M\ceil{2\varepsilon (1-1/p)\log_p L} \log_q p +o(M).
\end{align*}
Since  $2\varepsilon (1-1/p)<2\varepsilon$,  this redundancy is usually better than the Gilbert-Varshamov bound in Corollary~\ref{cor:LM-upbound-3}. When $(\kp,\km)=(1,0)$ and $p=2$, it almost meets the sphere-packing bound $M\varepsilon \log_q L+o(M)$.}

{

Finally, we would like to discuss the case where $\varepsilon=L$. We
note that by using $q$-ary deletion-correcting codes, one can easily
generalize Theorem~\ref{th:con-infepsilon} to obtain $q$-ary
$(s,t,\bullet)_{\bS}$-correcting codes of redundancy
\[(s+2t) L+ O(\log_q \log_q M).\]
This code can be used as an $(s,t,\bullet,\kp,\km)_{\bLM}$-correcting code.
In contrast, Corollary~\ref{cor:LM-upbound-3} shows the existence of such a code of  redundancy no more than
\[sL+2tL \log_q (\kp+\km+1)+(s+2t)\log_q M+O(1) .\]
Since $\log_q(\kp+\km+1)$ is less than one, in some cases this Gilbert-Varshamov bound is less than $(s+2t)L$. It is therefore an interesting question to find explicit constructions of codes with redundancy less than $(s+2t)L$. Besides, establishing a good lower bound on the redundancy of such codes is also an open problem.
}

{
We close  this section with a  description of the  quaternary case. Unlike the binary case, a quaternary sequence  produced by $n$ rounds of synthesis process is represented by a sequence \[\vd \circ \vr \eqdef ((d_1,r_1),(d_2,r_2),\ldots, (d_n,r_n)) \in (\Z_3\times \Z_q)^n,\]
where the sequence $\vd=(d_1,d_2,\ldots,d_n)$ represents the difference between the letters appended in the $(i-1)$th round  and $i$th round, or the difference between the letters in the initiator and in the first round (e.g., ref \cite{LeeKalGoeBolChur19}), and the sequence $\vr=(r_1,r_2,\ldots,r_n)$ still represents the run-lengths.

Our codeword is a subset $S=\set{\vd_1\circ \vr_1,\vd_2\circ\vr_2,\ldots, \vd_M \circ \vr_M}$ of  $(\Z_3\times \Z_q)^L$. The $(s,t,\varepsilon,\kp,\km)_{\bLM}$-DNA storage channel outputs a subset $S'$ of $S$, with at most $s$ sequences  lost and at most $t$   sequences  corrupted.  In  each erroneous sequence $\vd_i \circ \vr_i$, the sequence $\vd_i$ is preserved and at most $\varepsilon$ elements of   $\vr_i$ are corrupted  by the  $(\kp,\km)$-limited-magnitude errors.

In the following we briefly describe  how to generalize the constructions in Theorem~\ref{thm:conmod} and Lemma~\ref{lm:concatenation-LM} to yield codes of $(\Z_3\times \Z_q)^L$ with the same error-correcting capability. We do the encoding such that $(\vd_1,\vd_2,\ldots,\vd_M)$ is a codeword of an $s$-erasure-correcting code over $\F_{3^L}$, and $\set{\vr_1,\vr_2,\ldots,\vr_M}$ is a codeword of the  codes over sets in Theorem~\ref{thm:conmod} or Lemma~\ref{lm:concatenation-LM}. Note  that the constructions ensure that $\{\vr_1,\vr_2,\ldots,\vr_{M}\}$ is a block-code correcting $\varepsilon$ $(k_+,k_-)$-limited-magnitude errors. Assume that $S'=\set{\vd_1'\circ \vr_1',\vd_2'\circ \vr_2',\ldots,\vd_{M-s}'\circ \vr_{M-s}'}$. We can first recover the set $\{\vr_1,\vr_2,\ldots,\vr_{M}\}$ from $\set{ \vr_1',\vr_2',\ldots, \vr_{M-s}'}$ by using the decoding schemes in the proofs of Theorem~\ref{thm:conmod} or Lemma~\ref{lm:concatenation-LM}. Since $\{\vr_1,\vr_2,\ldots,\vr_{M}\}$ is an $(\varepsilon,k_+,k_-)$-error-correcting code,  by comparing the sequences $\vr_i'$ with the sequences $\vr_i$, we can  determine the ordering of the  received sequences $\vd_i'\circ \vr_i'$. In this way we actually determine the ordering of the $M-s$ surviving sequences $\vd_i$'s. Thus we can use the $s$-erasure-correcting code to recover the lost $s$ sequences.
}

\bibliographystyle{IEEEtranS}

\end{document}